# Generation of Entanglement Outside of the Light Cone

J.D. Franson

*University of Maryland, Baltimore County, Baltimore, MD 21250*

The Feynman propagator has nonzero values outside of the forward light cone. That does not allow messages to be transmitted faster than the speed of light, but it is shown here that it does allow entanglement and mutual information to be generated at space-like separated points. These effects can be interpreted as being due to the propagation of virtual photons outside of the light cone or as a transfer of pre-existing entanglement from the quantum vacuum. The differences between these two interpretations are discussed.

## 1. INTRODUCTION

Entanglement is one of the most fundamental and nonclassical aspects of quantum mechanics. It is shown here that entanglement, mutual information, and optical coherence can all be generated at two distant locations in less time than it would take for light to travel between them. These counterintuitive effects are possible because the probability amplitude to emit a photon at one location and annihilate it at another location is proportional to the Feynman propagator $D_F$ [1]. As noted by Feynman himself [2, 3], $D_F$ has nonzero values outside of the forward light cone, as illustrated in Fig. 1. It should be emphasized that this property of the Feynman propagator does not allow messages to be transmitted between space-like separated points, but it will be shown here that it does allow entanglement to be generated outside of the light cone.

The fact that the Feynman propagator is not confined to the light cone raises some issues with regard to causality that have previously been discussed by Feynman [2, 3], Hegerfeldt [4], and others [5-13]. Much of this discussion has been concerned with the probability that an excited atom at one location will emit a photon that is absorbed by a second, distant atom after some time interval $\Delta t$, as illustrated in Fig. 2. The probability $P_2$ that the second atom will be excited was first considered by Fermi [14], who made several unwarranted assumptions [15] and concluded that there were no effects at all outside of the forward light cone. Other authors [4-13, 15-20] considered the problem in more detail, some of whom [4, 20] concluded that there should be an increase in $P_2$ in apparent contradiction with causality. It was subsequently shown that $P_2$ cannot change outside the light cone when all possible effects are included [5-13], but that correlations between the states of the two atoms could be produced [5, 7, 11].

This paper generalizes the earlier results to show that entanglement can be generated outside of the light cone as well. It should be emphasized that the generation of correlations or entanglement in this way does not contradict the results of an earlier paper by Milonni et al [9], who considered the expectation values associated with the state of a distant atom and showed that causality is maintained. That paper did not consider the generation of correlations or entanglement between two distant atoms, which do not violate causality.

This paper and the references quoted above are all based on the usual perturbative approach to quantum optics or quantum electrodynamics, where these effects can be interpreted as being due to the propagation of virtual photons outside of the light cone. In a separate series of papers, Summers and Werner [21-25] used the more abstract techniques of algebraic quantum field theory [21-28] to show that Bell's inequality can be violated in the vacuum state of any quantum field theory that satisfies certain assumptions. In a similar manner, Reznik and his collaborators [29-33] have discussed the generation of entanglement outside of the light cone by scalar or Dirac fields, which they interpreted as being due to the transfer of pre-existing entanglement in the quantum vacuum to the atomic states. Thus the usual perturbative approach and the algebraic quantum field approach lead to very different interpretations of these effects.

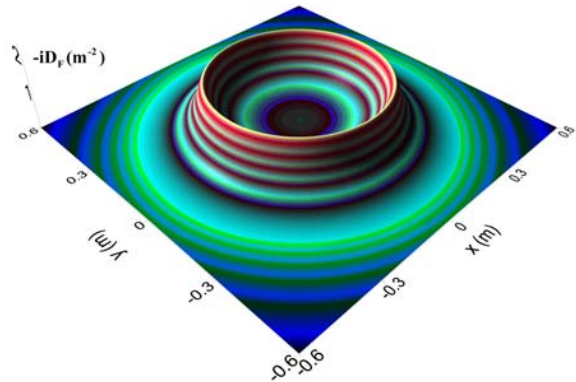

FIG. 1. A plot of the Feynman propagator $D_F$ in the xy plane, which is proportional to the probability amplitude to emit a photon at one location and annihilate it at another location. The plot shown here corresponds to a 1 ns delay after a photon was emitted at the origin. The colors of the contour levels were arbitrarily chosen to reflect the wave-like nature of a photon. The value of $D_F$ was truncated on the light cone (the yellow ring), where it diverges, but it can be seen that it has a nonzero value arbitrarily far outside the forward light cone as well.

Despite the long history of papers on this subject, there is still no consensus of opinion regarding the significance or interpretation of effects of this kind. The goal of this paper is to consider these issues in more detail and within the context of entanglement, quantum information theory, and the Feynman propagator. An attempt will be made to address all of the significant objections that have been raised in the past, such as the difficulty in localizing relativistic free particles, the use of the dipole approximation, and the virtual photon "cloud" associated with dressed atomic states. It is hoped that this will provide some additional insight into what is entailed by Fig. 1 as well as the difference between these two interpretations.

The remainder of the paper is organized as follows. In Section 2, perturbation theory and the Feynman propagator are used to calculate the probability amplitude to excite the second atom in the limit where $\Delta t$ is much smaller than the time required for light to travel between the two atoms. Those results are used in Section 3 to show that the two systems become entangled outside of the forward light cone, and that the entanglement of the two atoms can be increased using post-selection and entanglement concentration. These effects are compared with quantum teleportation and entanglement swapping in Section 4.

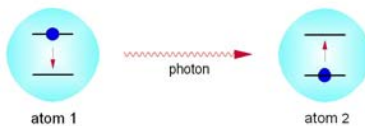

FIG. 2. Two distant atoms, one initially in its excited state and the other in its ground state. A photon emitted by atom 1 can be absorbed by atom 2 to produce a transition to its excited state. The probability amplitude for this process is determined by the Feynman propagator of Fig. 1 and is nonzero outside the forward light cone.

Section 5 considers the impact of these effects on special relativity, including a proof based on the commutation relations of the field operators that superluminal messages cannot be transmitted; it will be shown that the generation of entanglement and mutual information is not limited by that argument. The generation of optical coherence (as conventionally defined) outside the forward light cone is discussed in Section 6, and it is suggested that the definition of optical coherence may need to be reconsidered. The possibility of new types of quantum information protocols, including a quantum time capsule, is considered in section 7. The two interpretations mentioned above are contrasted and compared in Section 8, while a summary and conclusions are given in Section 9.

For simplicity, all of the calculations in the main text will be based on the dipole approximation. The effects of higher-order multipoles are included in Appendix A, where it is shown that they are negligible for the situation of interest here. The calculations in the main text consider only bare atomic states, which provide the most straightforward way to derive the effects of interest. Appendix B includes the effects of the virtual photons associated with the use of dressed atomic states. It is found that similar results are obtained in either the bare-state or dressed-state basis, which justifies the use of bare atomic states in the main text. These results do not depend on any relativistic properties of the atoms, but a covariant calculation using second-quantized Dirac operators is discussed in Appendix C nevertheless.

## 2. PERTURBATION THEORY CALCULATION

The effects of interest can be understood by considering two distant atoms labeled 1 and 2 and located at positions $\mathbf{x}_1$ and $\mathbf{x}_2$, as illustrated in Fig. 2. At the initial time $t = 0$, atom 1 is assumed to be in its first excited state $|E_1\rangle$ while atom 2 is in its ground state $|G_2\rangle$, with the field in its vacuum state (no photons present). The system is then allowed to evolve for a time interval $\Delta t$, during which there is some probability amplitude that atom 1 may emit a photon and make a transition back to its ground state $|G_1\rangle$ while atom 2 absorbs the photon and makes a transition to its excited state $|E_2\rangle$. Since the Feynman propagator is nonzero outside the forward light cone, the question is whether or not this process could produce a state of the form

$$|\psi\rangle = a|E_1\rangle|G_2\rangle + b|G_1\rangle|E_2\rangle + \gamma|\phi_\perp\rangle \tag{1}$$

even if $|\mathbf{x}_2 - \mathbf{x}_1| > c\Delta t$. Here the state $|\phi_\perp\rangle$ is orthogonal to the other two states and it includes the possibility that atom 1 may have emitted a photon that was not absorbed by atom 2, for example.

Another scenario would be to assume that the excited states $|E_1\rangle$ and $|E_2\rangle$ correspond to metastable states with zero dipole moment, such as the 2S state of hydrogen. In that case, atom 1 can be assumed to have been in state $|E_1\rangle$ for times $t < 0$ with negligible interaction with the electromagnetic field, and similarly for atom 2 in state $|G_2\rangle$. A dipole moment could then be produced by applying an external electric field to both atoms over the time interval $\Delta t$, as in the Lamb shift experiments. To the extent that the interaction with the field is negligible outside of the time interval $\Delta t$, the analysis of this situation is the same as that of the original example of Fig. 2. This example has several advantages when dressed atomic states are considered and it will be discussed in more detail in Appendix B.

The probability amplitude $b$ to find the system in the state $|G_1\rangle|E_2\rangle$ with no photons present [34] at time $\Delta t$ can be calculated in a straightforward way using commutators and the Feynman propagator. From second-order perturbation theory, which is equivalent [35] to the use of Wick's theorem in scattering (S-matrix) calculations, the change in the state of the system at time $\Delta t$ is given by

$$|\psi(\Delta t)\rangle = \frac{1}{(i\hbar)^2} \int_0^{\Delta t} dt' \int_0^{t'} dt'' \hat{H}'(t') \hat{H}'(t'') |\psi_0\rangle. \qquad (2)$$

The interaction Hamiltonian $\hat{H}'(t)$ is given [36] by

$$\hat{H}'(t) = \frac{e}{c} \int d^3\mathbf{r} \, \hat{\mathbf{j}}(\mathbf{r},t) \cdot \hat{\mathbf{A}}(\mathbf{r},t) \qquad (3)$$

where $\mathbf{j}(\mathbf{r},t)$ is the current operator, $\hat{\mathbf{A}}(\mathbf{r},t)$ is the vector potential operator, and $-e$ is the charge of the electron. The minimum coupling Hamiltonian of Eq. (3) has the advantage of being manifestly covariant in the Lorentz gauge (see appendix C) and is used in quantum electrodynamics. Similar results for the correlations between the atoms were also obtained by Power and Thirunamachandran [11] using the $\mathbf{r} \cdot \mathbf{E}$ form of the Hamiltonian.

In the limit of small atomic dimensions (dipole approximation), the vector potential can be evaluated at the centers of the atoms and the atomic matrix elements of $\mathbf{j}$ reduce to $-iE_A\mathbf{d}/\hbar$, where $\mathbf{d} = \langle G|\mathbf{x}|E \rangle$ is the atomic dipole moment, $E_A$ is the energy of the excited atomic states, and $\hbar$ is Planck's constant divided by $2\pi$ [36]. (The contribution from higher-order multipoles is negligible, as shown in Appendix A.) The projection of equation (2) onto the state $|G_1\rangle|E_2\rangle|0\rangle$ then gives

$$b = \frac{1}{(i\hbar)^2} \left(\frac{edE_A}{\hbar c}\right)^2 \int_0^{\Delta t} dt' \int_0^{t'} dt'' e^{-iE_A(t''-t)/\hbar} \langle 0|\hat{A}_x(\mathbf{x}_2,t')\hat{A}_x(\mathbf{x}_1,t'')|0\rangle \qquad (4)$$

where $|0\rangle$ is the vacuum state of the field with no photons and it has been assumed for simplicity that the dipole moment is perpendicular to $\mathbf{x}_2 - \mathbf{x}_1$ and along the x-axis.

As usual [37], we can write the vector potential as the sum of its positive and negative frequency components

$$\hat{A}_x(\mathbf{x},t) = \hat{A}_x^{(+)}(\mathbf{x},t) + \hat{A}_x^{(-)}(\mathbf{x},t) \qquad (5)$$

where $\hat{A}_x^{(-)}(\mathbf{x},t)$ creates a photon and $\hat{A}_x^{(+)}(\mathbf{x},t)$ annihilates a photon. Only the product $\hat{A}_x^{(+)}(\mathbf{x}_2,t')\hat{A}_x^{(-)}(\mathbf{x}_1,t'')$ contributes to the matrix element $M$ in Eq. (4) and

$$M \equiv \langle 0|\hat{A}_x(\mathbf{x}_2,t')\hat{A}_x(\mathbf{x}_1,t'')|0\rangle = \langle 0|\hat{A}_x^{(+)}(\mathbf{x}_2,t')\hat{A}_x^{(-)}(\mathbf{x}_1,t'')|0\rangle. \qquad (6)$$

Equation (6) can be simplified using the commutator of the field operators to give

$$M = \langle 0|\hat{A}_x^{(-)}(\mathbf{x}_1,t'')\hat{A}_x^{(+)}(\mathbf{x}_2,t') + [\hat{A}_x^{(+)}(\mathbf{x}_2,t'),\hat{A}_x^{(-)}(\mathbf{x}_1,t'')]|0\rangle$$
$$M = \langle 0|[\hat{A}_x^{(+)}(\mathbf{x}_2,t'),\hat{A}_x^{(-)}(\mathbf{x}_1,t'')]|0\rangle. \qquad (7)$$

The commutator has a simple form in the Lorentz gauge [37], where

$$[\hat{A}_x^{(+)}(\mathbf{x}_2,t'),\hat{A}_x^{(-)}(\mathbf{x}_1,t'')] = -ic\hbar D_F(\mathbf{x}_2 - \mathbf{x}_1, t' - t'') \qquad (8)$$

for $t' > t''$ [where $\theta(t'-t'')=1$]. Combining Eqs. (4), (7), and (8) gives

$$b = -\frac{1}{i\hbar^2}\left(\frac{edE_A}{\hbar c}\right)^2 \int\limits_0^{\Delta t} dt' \int\limits_0^{t'} dt'' e^{-iE_A(t'-t'')/\hbar} D_F(\mathbf{x_2}-\mathbf{x_1}, t'-t'') \qquad (9)$$

For a massless particle such as a photon, $D_F$ has the explicit value [38, 39]

$$D_F(x_2-x_1) = -\frac{1}{4i\pi^2}\frac{1}{(x_2-x_1)^2 - i\varepsilon}$$

$$= -\frac{1}{4i\pi^2}\frac{1}{\left|\mathbf{x_2}-\mathbf{x_1}\right|^2 - c^2(t'-t'')^2 - i\varepsilon} \qquad (10)$$

Here $\varepsilon$ is an infinitesimally small quantity that determines the value of any integrals and we have used the 4-vector notation $x = (\mathbf{x}, ict)$, which demonstrates the covariant nature of the propagator. We will consider a space-like separation with $r = \left|\mathbf{x_2}-\mathbf{x_1}\right| >> c\Delta t$, which gives $D_F = -1/4i\pi^2 r^2$. This approximation is valid far outside of the light cone and it is equivalent to neglecting terms that are of order $(\Delta t / r)^2$ smaller than the remaining terms, as can be seen from a power series expansion of Eq. (10).

The two integrals over time in Eq. (9) can then be evaluated to give

$$b = -\frac{\alpha}{4\pi^2}\frac{d^2}{r^2}\left(i\omega_A\Delta t + 1 - e^{i\omega_A\Delta t}\right) \qquad (11)$$

Here $\alpha = e^2/\hbar c$ is the fine structure constant and $\omega_A = E_A/\hbar$ is the resonant frequency of the atomic transition.

Eq. (11) corresponds to the probability amplitude for atom 1 to emit a photon that is absorbed by atom 2. Although it may seem counterintuitive, there is also a probability amplitude for atom 2 to emit a photon and make a transition to its excited state while atom 1 absorbs the photon and makes a transition to its ground state, since a virtual process of that kind need not conserve energy in the intermediate state. This probability amplitude can be calculated in the same way as that leading to Eq. (11) and the total probability amplitude from both processes is given by

$$b = -\frac{\alpha}{2\pi^2}\frac{d^2}{r^2}\left[1 - \cos(\omega_0\Delta t)\right]. \qquad (12)$$

Eq. (12) shows that there is a probability amplitude for the two atoms to exchange a photon even though they are space-like separated, and this entangles the two systems as discussed in the next section. It can be seen that the effects are only appreciable for atomic separations that are at most a few orders of magnitude larger than $d$. As a result, the predicted correlations are experimentally observable in principle but an actual experimental test would be difficult.

It should be emphasized once again that these results do not contradict an earlier paper by Milonni et al. [9], which only considered the expectation value of the state of atom 2 and not the correlations between the two atomic states. Commutator techniques are used in section 5 to give a more general proof that the expectation values associated with atom 2 cannot change outside of the light cone, as would be expected from causality.

### 3. ENTANGLEMENT GENERATION AND CONCENTRATION

By definition, an entangled state is any quantum state that cannot be written as the product of two or more single-particle states. Eq. (1) does not appear to be factorable, but the situation is complicated by the presence of the $|\phi_\perp\rangle$ term and a more careful analysis is required in order to demonstrate that the state $|\psi\rangle$ actually is entangled.

If we could simply ignore the $|\phi_\perp\rangle$ term we would have the entangled state $|\psi'\rangle = a|E_1\rangle|G_2\rangle + b|G_1\rangle|E_2\rangle$, where the coefficients $a$ and $b$ have been normalized to give unit probability. That term cannot be ignored, however, and one way to include its effects is to ask whether or not the state $|\psi\rangle$ of Eq. (1) can be written in the form

$$|\psi\rangle = |\Psi_1\rangle|\Psi_2\rangle \tag{13}$$

where $|\Psi_1\rangle$ represents the most general form of the system $S_1$ consisting of atom 1 and any localized field associated with it:

$$\begin{aligned} |\Psi_1\rangle = |G_1\rangle \sum_{\{n_i\}} c_1(\{n_i\})\left(\hat{a}_1^\dagger\right)^{n_1} \cdots \left(\hat{a}_N^\dagger\right)^{n_N} |0\rangle \\ + |E_1\rangle \sum_{\{n_i\}} d_1(\{n_i\})\left(\hat{a}_1^\dagger\right)^{n_1} \cdots \left(\hat{a}_N^\dagger\right)^{n_N} |0\rangle. \end{aligned} \tag{14}$$

Here $\hat{a}_i^\dagger$ creates a photon with wave vector $\mathbf{k_i}$ and $c_1(\{n_i\})$ and $d_1(\{n_i\})$ are arbitrary complex coefficients. A similar expression exists for the most general form of the system $S_2$ consisting of atom 2 and any field associated with it. Taking the projection onto the vacuum state gives

$$\begin{aligned} |0\rangle\langle 0|\psi\rangle = |0\rangle\Big[ c_1(0)|G_1\rangle + d_1(0)|E_1\rangle \Big] \\ \times \Big[ c_2(0)|G_2\rangle + d_2(0)|E_2\rangle \Big] = |\psi'\rangle. \end{aligned} \tag{15}$$

But $|\psi'\rangle$ is entangled and cannot be written in this form [40], which implies that Eq. (13) cannot hold. Thus the two systems $S_1$ and $S_2$ must be entangled as well, even without the post-selection process described below. This result corresponds to the use of bare atomic states, but a more general result is derived in Appendix B for dressed atomic states.

Most applications of quantum information would use only the atoms independent of the state of the field, and an average (trace) over the field states would then have to be taken. Although $|\psi\rangle$ itself is a pure state, the trace would produce a mixed state with a density matrix $\hat{\rho}$ that would no longer be entangled. It can be shown [11], however, that the atoms are still correlated if a trace is taken over the field states.

Entangled states of the atoms alone could be produced, at least in principle, by using an array of single-photon detectors to determine whether or not any photons were present in the final state.

It will be assumed that suitable detectors are located throughout the region surrounding the two atoms and that they are all turned on shortly after time $\Delta t$, with a response time that is comparable to $\Delta t$. If no photons are detected, the system would be projected into the entangled state $|\psi'\rangle$ at a time $\Delta t' \sim \Delta t$, which is outside the forward light cone.

The use of an array of detectors would allow the post-selection process illustrated in Fig. 3a, where the event is rejected if a photon is found after time $\Delta t$, as indicated by the red X in the figure. In principle, this post-selection process could be performed using a large number of pairs of atoms to generate a smaller number $N$ of pairs of atoms in entangled states. In order to actually use this entanglement in a quantum information protocol, it would be necessary to wait for a much longer time $\Delta T \geq |\mathbf{x}_2 - \mathbf{x}_1|/c$ in order to distribute the results from the detectors to all of the relevant locations. Nevertheless, the entangled states were created at time $\Delta t'$ as recorded by the detectors.

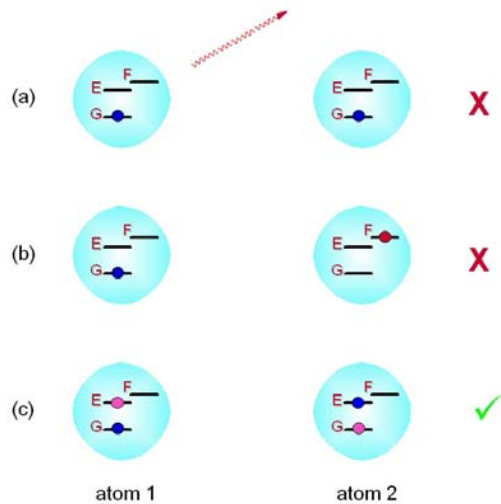

FIG. 3. Generation of maximally entangled atoms using post-selection. (a) If a photon is detected after time $\Delta t$, the event is rejected. (b) If atom 2 is found in state F, the event is also rejected as described in more detail in Fig. 4. (c) If no photon is found and atom 2 is not in state F, the two atoms are known to have been maximally entangled outside of the light cone. An entangled state of the atoms and their associated fields will be generated outside the light cone without any need for post-selection.

Most quantum information protocols would also require that the entanglement of the two atoms be nearly perfect (maximal entanglement). This could be achieved (in principle) using entanglement concentration [41], such as the protocol illustrated in Fig. 4. Here the ground state $|G_2\rangle$ of atom 2 is coupled with a laser pulse to a third atomic level $|F_2\rangle$. By adjusting the intensity of the laser pulse, it is possible to transfer any desired amount of probability amplitude from $|G_2\rangle$ to $|F_2\rangle$ and to convert the post-selected state $|\psi'\rangle$ into the state $|\psi''\rangle$ given by

$$|\psi''\rangle = b|E_1\rangle|G_2\rangle + b|G_1\rangle|E_2\rangle + \gamma'|E_1\rangle|F_2\rangle \qquad (16)$$

where $\gamma'$ is a complex constant of no interest. A measurement is then performed to determine whether or not atom 2 is in state $|F_2\rangle$. If it is, then the event is discarded as illustrated by the red X in Fig. 3b. If not, then the system will be projected into the maximally entangled state $\left(|E_1\rangle|G_2\rangle + |G_1\rangle|E_2\rangle\right)/\sqrt{2}$. A protocol of this kind can be used to convert $N$ pairs of partially entangled atoms into $N' < N$ pairs of maximally entangled Bell states.

It should be noted that the laser pulse and measurements can be completed outside the light cone in a small time interval after time $\Delta t$. Similar techniques have been proposed [42] for use in Zeno quantum logic gates [43], and Reznik et al. [29-33] have considered the use of entanglement distillation with scalar fields.

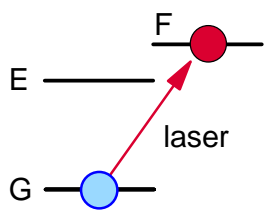

FIG. 4. Concentration of entanglement in which $N$ partially-entangled pairs of atoms are converted into a smaller number $N'$ of maximally entangled atoms. A laser beam is used to partially excite atom 2 from its ground state to state F. A pair of atoms is subsequently rejected if a measurement shows that atom 2 is in state F. This process creates a balanced superposition of states $|E_1\rangle|G_2\rangle$ and $|G_1\rangle|E_2\rangle$.

To summarize this section, it has been shown that an entangled state of the atoms and their associated fields will be generated outside the light cone without any need for post-selection. A maximally entangled state of the two atoms alone can be prepared using post-selection and entanglement concentration, where the entangled states are known to have been produced within time $\Delta t' < |\mathbf{x}_2 - \mathbf{x}_1|/c$. The atoms are still correlated but not entangled if an average over the field states is taken instead.

## 4. COMPARISON WITH QUANTUM TELEPORTATION AND ENTANGLEMENT SWAPPING

The post-selection processes described above could be used to prepare $N'$ pairs of maximally entangled atoms. Entanglement between two distant systems can also be prepared using quantum teleportation [44] or entanglement swapping [45], and it may be useful to discuss the fundamental differences between these techniques.

In quantum teleportation, two experimenters A (Alice) and B (Bob) would like to transport an unknown quantum state from Alice to Bob. It is assumed that they already share an entangled pair of particles, such as a pair of photons that were generated in an entangled state and then propagated through optical fibers to Alice and Bob. Alice performs a Bell state measurement between her unknown state and her member of the entangled pair, which produces two bits of classical information. Once the classical information is transmitted to Bob, it can be used to regenerate the unknown quantum state by applying a suitable transformation to his member of the entangled pair of particles.

Although quantum teleportation is a remarkable and useful process, it is apparent that the information needed to recreate Alice's unknown state is transmitted in the form of classical information from Alice to Bob at ordinary velocities. This is very different from the situation described above, where the entanglement was created in a time interval $\Delta t'$ outside of the forward light cone and the classical information is used only to determine which events were successful; no corrections are applied to the atoms based on that information.

In entanglement swapping [45], a pair of entangled particles is created locally at Alice and Bob. One member of each pair is then transmitted at ordinary velocities to a central location, where a Bell state measurement is made. If we post-select on the results of the measurement, the two distant particles will be left in an entangled state. But in this case the entangled state is not generated until sufficient time has elapsed for two of the particles to have traveled to the same location. In addition, there is no physical interaction of the distant particles via virtual photons as there is in the situation of interest here.

## 5. SPECIAL RELATIVITY AND CAUSALITY

The generation of entanglement outside of the forward light cone may seem counterintuitive, but it does not allow messages to be transmitted faster than the speed of light. These effects are thus consistent with special relativity to that extent, although they raise some questions regarding the postulates of relativity theory.

Commutator techniques can be used to provide a simple proof that there can be no net change in the probability $P_2$ of finding atom 2 in its excited state. In analogy with the usual perturbation theory expression for the state vector in Eq. (2), the expectation value of any observable quantity corresponding to an operator $\hat{Q}$ at time t is given in the interaction picture [46] by

$$
\langle \psi(t) | \hat{Q} | \psi(t) \rangle = \langle \psi_0 | \hat{Q} | \psi_0 \rangle + \frac{1}{i\hbar} \int_0^t dt' \langle \psi_0 | \left[ \hat{Q}, \hat{H}'(t') \right] | \psi_0 \rangle
$$
$$
+ \frac{1}{(i\hbar)^2} \int_0^t dt' \int_0^{t'} dt'' \langle \psi_0 | \left[ \left[ \hat{Q}, \hat{H}'(t') \right], \hat{H}'(t'') \right] | \psi_0 \rangle + ...
$$

(17)

Here the interaction Hamiltonian $\hat{H}'(t)$ involves the free-field (Heisenberg picture) operators $\hat{\mathbf{A}}(\mathbf{x}_1, t)$ and $\hat{\mathbf{A}}(\mathbf{x}_2, t)$, which satisfy the usual commutation relations. The probability of finding atom 2 in its excited state is given by the expectation value of the projection operator $\hat{p}_2 = | E_2 \rangle \langle E_2 |$. Since this operator and the corresponding field operator commute with the Hamiltonian of atom 1 and the field at that location, it follows immediately that $P_2$ cannot depend on the state of atom 1 outside the light cone. On the other hand, the correlated probability described by the operator $\hat{p}_{12} = | E_2 \rangle | G_1 \rangle \langle G_1 | \langle E_2 |$ does not commute with atom 1 and need not be zero outside of the forward light cone.

The fact that there is no net change in $P_2$ can be understood from the fact that there are other processes that must be included as well. For example, energy conservation does not strictly apply for a finite time interval $\Delta t$, where the uncertainty relation $\Delta E \, \Delta t \geq \hbar / 2$ holds. As a result, atom 2 could emit a photon and make a transition to its excited state even if atom 1 were not present. The probability to find atom 2 in its excited state with a photon present is reduced if atom 1 can absorb the photon, and this tends to cancel the increase in the probability $\Delta P_2$ calculated above. A complete calculation of $P_2$ must include all fourth-order terms and gives no dependence on the initial state of atom 1 [5-13]. Thus the exchange of a photon can produce correlations and entanglement between the two atoms, but those correlations cannot be controlled by an experimenter to send superluminal messages between the two locations.

The postulates of special relativity are that (i) the laws of nature are the same in all inertial reference frames and (ii) the speed of light is a constant independent of its source. In order to investigate the impact on postulate (i), consider another reference frame R' that is moving at a velocity $\mathbf{v}$ with respect to the original reference frame R in which the two atoms are allowed to interact with the field over the same time interval $\Delta t$. (Here we assume metastable excited states with the dipole interaction controlled by an external electric field.) Since the events of interest are space-like separated, it is possible to choose $\mathbf{v}$ in such a way that atom 1 is allowed to interact over a time interval $\Delta t_1$ that does not overlap with the time interval $\Delta t_2$ over which atom 2 is allowed to interact. Moreover, the interaction at atom 1 can occur before or after the interaction at atom 2 depending on the choice of $\mathbf{v}$. It can be shown that the same entanglement is produced in any inertial reference frame, in agreement with postulate (i). This result is ensured by the fact that perturbation theory can be put into an invariant form [47].

Although the same entanglement is produced in any reference frame, the photon can only be emitted by atom 1 and travel to atom 2 in some reference frames, whereas it can only travel from atom 2 to atom 1 in other reference frames. As a result, any causal interpretation of this process would depend on the choice of reference frame. In that sense, there is some similarity to the collapse or reduction of the wave function in experiments based on Bell's inequality [48].

The second postulate assumes that the speed of light is a constant in all reference frames regardless of its source. Special relativity was proposed in the context of classical electromagnetism where the speed of light has a well-defined value, and quantum electrodynamics did not exist at the time. The effects described above correspond to the generation of entanglement outside of the forward light cone by the exchange of virtual photons, which raises some possible issues regarding the definition of the speed of light. For example, do individual photons travel faster than the classical speed of light during the time that they are being exchanged? Are the photons really being exchanged if we are not allowed to detect them in the process? And what is the nature of the photons if their direction of travel depends on the choice of reference frame?

Similar questions arise in the interpretation of Feynman diagrams, which consist of space-time points (vertices) where two or more particles interact. The particles propagate between the vertices as described by the Feynman propagator, which connects points that are space-like separated with a small but nonzero probability amplitude. One might ask once again whether or not the particles travel faster than the speed of light over such trajectories. As Feynman put it [3], "possible trajectories are not limited to regions within the light cone" but "in reality, not much of the $t - |\mathbf{x}|$ space outside the light cone is accessible".

It seems apparent that these effects are due to the properties of the Feynman propagator and that the atoms can be treated nonrelativistically. Nevertheless, a covariant calculation using second-quantized Dirac operators is outlined in Appendix C.

## 6. OPTICAL COHERENCE

In the conventional theory of single-photon detection developed by R.J. Glauber [49], the probability $P_d$ of detecting a single photon at position $\mathbf{x}$ and time $t$ is given by

$$P_d = \eta \left\langle \hat{E}^{(-)}(\mathbf{x},t)\hat{E}^{(+)}(\mathbf{x},t) \right\rangle \tag{18}$$

Here $\eta$ is a constant related to the detector efficiency and time window. This expression can be evaluated in the same way as $\Delta P_2$ using the Feynman propagator, with the result that $P_d$ would be nonzero outside of the forward light cone, in apparent violation of causality. However, Eq. (18) is based on an approximation [8, 9, 50] that neglects the possibility that the detector may emit a photon as well as absorb one (the rotating wave approximation). As described above, that process will cancel the more intuitive one and the actual detection probability is unchanged outside of the forward light cone, as can be shown using Eq. (17) or other methods.

The concept of higher-order optical coherence plays an important role in quantum optics. If we assume that atom 1 was initially in its ground state and then excited with a short laser pulse at time $t = 0$, we can ask whether or not there is any coherence between the remaining laser pulse and the field in the vicinity of atom 2. In analogy with Eq. (18), the second-order optical coherence at two locations is defined [49] as

$$g^{(2)}(\mathbf{x_1},t_1;\mathbf{x_2},t_2) = \frac{\left\langle \hat{E}^{(-)}(\mathbf{x_1},t_1)\hat{E}^{(-)}(\mathbf{x_2},t_2)\hat{E}^{(+)}(\mathbf{x_2},t_2)\hat{E}^{(+)}(\mathbf{x_1},t_1) \right\rangle}{\left\langle \hat{E}^{(-)}(\mathbf{x_1},t_1)\hat{E}^{(+)}(\mathbf{x_1},t_1) \right\rangle \left\langle \hat{E}^{(-)}(\mathbf{x_2},t_2)\hat{E}^{(+)}(\mathbf{x_2},t_2) \right\rangle} \tag{19}$$

If we accept this definition, then it can be shown that second-order coherence would be generated between the laser pulse used to excite atom 1 and the field at a space-like separated point. Once again, this is a non-physical result due to the use of the rotating-wave approximation, and it suggests that the definition of optical coherence may need to be reconsidered [9,50].

## 7. QUANTUM INFORMATION PROTOCOLS

The nonclassical nature of quantum information allows a number of potentially useful practical applications, such as quantum computing and quantum key distribution. The ability to generate entanglement outside of the forward light cone would, in principle, allow the possibility of additional protocols that would not be possible otherwise. As a practical matter, the rate at which the entanglement can be generated would be too small to be of any real use, but it is interesting to consider protocols of this kind nonetheless.

One example of a quantum protocol of this kind is the quantum time capsule. In a conventional time capsule, information and artifacts describing the current environment are sealed into a capsule that is to be left unopened for some length of time, say 100 years. Classically, there is no way to ensure that the capsule might not be opened sooner than intended. This would be a potential problem, for example, if a well-known politician sealed his or her notes and memoirs in a time capsule with the understanding that the material would remain unavailable until long after his or her death.

A quantum-mechanical version of a time capsule that cannot be prematurely opened could (in principle) be implemented if entanglement were generated between two distant locations, say 100 light years apart, as illustrated in Fig. 5. Information could be encoded in the qubits at one location by taking the XOR (exclusive OR operation) between the classical information and the entangled qubits. The classical information could then be destroyed, and it could not be recreated from the qubits at the first location since they have random values [51]. The information could only be retrieved by bringing together the entangled pairs of atoms from the two distant locations and comparing the qubits (a second XOR operation), which could only be done after 100 years in this example. Generating the entanglement at a distance eliminates the possibility of cheating that would occur if the entangled pairs were generated at one location and then allegedly separated by a large distance. Although protocols of this kind are not practical, they do illustrate the fact that the entanglement generated in this way can have unique properties.

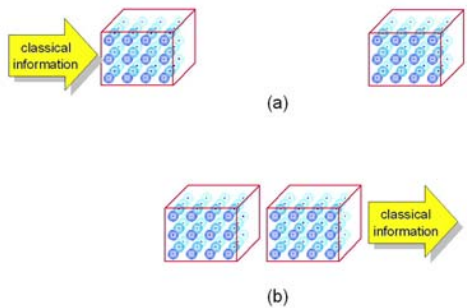

FIG. 5. Implementation of a quantum time capsule whose contents cannot be accessed for a specified period of time. (a) Two distant quantum memories contain a large number of atoms that are pair-wise entangled. Classical information is stored in one of the quantum memories by taking the XOR of the classical bits with the qubits in the memory and destroying the classical information. (b) The information can only be retrieved when the two quantum memories are brought together at less than or equal to the speed of light and a second XOR is performed.

It may also be useful to note that these effects allow mutual information as conventionally defined to be established outside of the light cone. Suppose that $N'$ pairs of maximally entangled qubits are generated at two space-like separated locations as described above. If the values of the qubits are simply measured, then the classical mutual information is defined by

$$I(X;Y) = \sum_{y \in Y} \sum_{x \in X} p(x,y) \log_2 \frac{p(x,y)}{p(x)p(y)} \tag{20}$$

where $p(x,y)$ is the joint probability distribution of the variables $X$ and $Y$ and $p(x)$ and $p(y)$ are their marginal probability distributions. Since the resulting measurements are totally correlated, the process yields $N'$ bits of mutual information that were generated outside of the forward light cone.

## 8. INTERPRETATION OF RESULTS

Effects similar to those described here were independently predicted based on two different approaches. Starting from the work of Fermi [14] and Hegerfeldt [4], the usual perturbative approach to quantum optics or quantum field theory led to the realization that correlations could be generated between two distant atoms. Here that approach was generalized to show that entanglement and mutual information could be generated outside the light cone as well. This approach involves the exchange of virtual photons and the Feynman propagator (or equivalent calculations). Similar effects were also predicted using algebraic quantum field theory, which leads to a different interpretation that does not involve virtual photons [21-33].

Electromagnetic interactions are generally viewed as being produced by the exchange of real or virtual photons, which is consistent with the usual perturbative treatment of quantum electrodynamics. The photons are assumed to carry momentum and energy and that is responsible for the force between two charges, for example. In that case, the most natural interpretation of the results obtained here would be to assume that the entanglement is generated by the propagation of virtual photons outside the forward light, as suggested by Fig. 1 and substantiated by the calculations of Section 2. This interpretation is complicated by the fact that the photons cannot be directly observed without destroying the effects of interest, while their direction of travel depends on the choice of reference frame, as discussed in Section 5.

Summers and Werner independently used the more abstract techniques of algebraic quantum field theory to show that Bell's inequality is violated by the field operators in the vacuum state of a quantum field [21-25]. Later work in algebraic field theory showed that the vacuum state of the field is entangled provided that the theory satisfies certain assumptions [26-33]. This has prompted an interpretation by Reznik and his collaborators [29-33] in which effects similar to those of interest here (but for a scalar field) were assumed to be due to a transfer of pre-existing entanglement from the quantum vacuum to the atoms, with no requirement for any transfer of information outside of the light cone.

The assertion that the quantum vacuum is entangled may seem surprising within the context of quantum optics. Whether a system is entangled or not depends on the choice of basis vectors. In the usual plane-wave number-state basis used in quantum optics, the quantum vacuum corresponds to the product of a large number of ground states of independent harmonic oscillators, which is not an entangled state. But the plane waves do not correspond to spatially separated systems and they do not form a suitable basis for a discussion of entanglement. In algebraic quantum field theory, the states of interest are localized to two or more separated regions and the quantum vacuum corresponds to an entangled state in that basis.

Regardless of whether the quantum vacuum is entangled or not, the vacuum fluctuations at two different locations are correlated. For example, it follows from Eq. (8) that

$$\langle 0 | \hat{A}_x(\mathbf{x}_2, t') \hat{A}_x(\mathbf{x}_1, t'') | 0 \rangle = -ic\hbar D_F(\mathbf{x}_2 - \mathbf{x}_1, t' - t'') \tag{22}$$

which shows that the field operators at two different locations are correlated, as qualitatively indicated in Fig. 6a. The algebraic quantum field approach suggests that these local fluctuations may be responsible for the change in the state of the two atoms without any requirement for a transfer of information outside of the light cone.

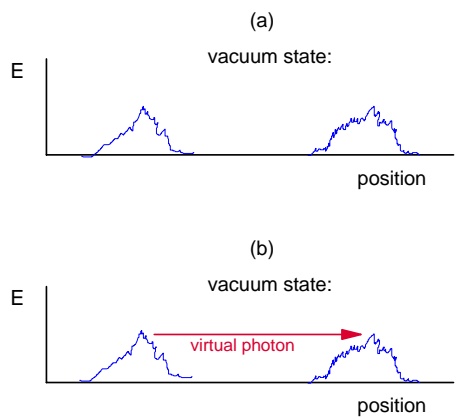

FIG. 6.  (a)  Correlations between the electric fields E due to vacuum fluctuations at two distant locations, which could produce correlated changes in the states of two atoms.    (b)  The state of an atom cannot change without a change in the state of the field, which corresponds to the emission or absorption of virtual photons.

But it seems to me that there is something missing here, namely the virtual photons illustrated in Fig. 6b [53].  If a test particle is placed in the electromagnetic field in its vacuum state, then conservation of momentum does not allow a change in the momentum of the test particle unless the state of the field changes as well.  Such a change in the field corresponds to the emission or absorption of a virtual photon.  The form of the field operators also shows that there can be no change in the state of the atoms unless there is a change in the state of the field corresponding to the emission or absorption of a virtual photon.  This suggests that the quantum vacuum alone is not sufficient to produce the effects of interest here, and that the emission and absorption of virtual photons is required.

The correlations between the field operators can be further understood if we insert a complete set of basis states in the left-hand side of Eq. (22).  Since the vector potential only couples the vacuum to single-photon states, this can be written as

$$\langle 0|\hat{A}_x(\mathbf{x_2},t')\hat{A}_x(\mathbf{x_1},t'')|0\rangle = \sum_k \langle 0|\hat{A}_x(\mathbf{x_2},t')|k\rangle\langle k|\hat{A}_x(\mathbf{x_1},t'')|0\rangle \tag{23}$$

where $|k\rangle$ is a single-photon state with wave vector $k$ .  In the presence of two test charges, each of the matrix elements on the right-hand side of the equation correspond to the emission or absorption of a photon, whose propagation is described by the Feynman propagator as before.  Eq. (23) suggests that the correlations between the vacuum fluctuations at two locations are maintained by the exchange of virtual photons, which is consistent with the fact that the states of the two atoms cannot change in a correlated way without such an exchange.

In contrast, the algebraic quantum field theory approach considers the correlations between the field operators without considering their origin.  As a result, the propagation of virtual photons does not enter into those discussions.  Unless we are willing to abandon the role of photons in electromagnetic interactions, the discussion above shows that the effects of interest here are due to the propagation of virtual photons outside of the forward light cone.

## 9.  SUMMARY AND CONCLUSIONS

It has been shown that entanglement, optical coherence, and mutual information can all be generated between two space-like separated points.  An entangled state of the atoms and the field can be generated outside the light cone without any need for post-selection, while the latter can be used to identify pairs of atoms that were maximally entangled after an arbitrarily short amount of time.

The analysis in the main text was based on the use of bare atomic states, but similar results are obtained in the dressed-state basis in Appendix B, where it is shown that the entanglement generated during the time interval $\Delta t$ is independent of the "cloud" of virtual photons that may have existed before that time.  This justifies the use of bare states in the main text.

It has been suggested that these results do not demonstrate any "real" entanglement because useful results can only be obtained using post-selection.  But Eq. (15) clearly shows that the two systems (including the field) become entangled without any post-selection or entanglement concentration; a generalization of this result to dressed atomic states is given in Appendix B.

Even in the case where post-selection is used, no corrections need be applied to the atoms and they must have been entangled during a time interval $\Delta t' < |\mathbf{x}_2 - \mathbf{x}_1| / c$, as recorded by the photon detectors.

These counterintuitive effects are due to the fact that the Feynman propagator has nonzero values outside of the forward light cone. Hegerfeldt [52] has noted that this is an unavoidable feature of any quantum field theory where the energies of the particles are bounded from below. If electromagnetic interactions are viewed as being produced by the exchange of virtual photons that carry energy and momentum, then the results obtained here can be interpreted as being due to the propagation of virtual photons outside of the light cone. Whether such effects should be considered to be superluminal is a semantic issue, since the photons cannot be directly observed and only the resulting correlations between the atoms can be measured.

An alternative interpretation has been suggested based on algebraic quantum field theory, where the quantum vacuum is considered to be an entangled state [21-33]. In that case, the effects of interest here could be interpreted instead as being due to the transfer of entanglement from the quantum vacuum to the atoms without the need for an exchange of particles or information outside of the forward light cone. Although that is perhaps a valid interpretation, it does not explain how the correlations in the vacuum fluctuations are produced or maintained. My personal preference would be to retain the usual role of virtual photons in quantum electrodynamics.

Regardless of the interpretation, a number of questions remain [53]: Are there any feasible experiments to test effects of this kind outside of the light cone? If an experiment were to be performed, would these effects be observed? And are there any alternative theories that do not have this property? Further research of this kind appears to be a natural extension of the earlier work on Bell's inequality.

## ACKNOWLEDGEMENTS

I am grateful to B.C. Jacobs, T.B. Pittman, and M.H. Rubin for their comments on the manuscript.

## APPENDIX A: DIPOLE APPROXIMATION

Eq. (4) in the text is based on the dipole approximation, which is valid when the characteristic dimensions of the atom are much smaller than the wavelength of the photons. Since the wavelength does not explicitly appear in the commutator approach used here, it may be useful to verify that the dipole approximation is valid for the situation of interest. It will be shown that the contribution from the higher-order multipoles is negligible for $r \gg c\Delta t$ and the only significant contribution is from the dipole term. The basic approach will be to evaluate the commutator of the field operators before the atomic matrix elements are calculated.

Combining Eqs. (2) and (3) in the text gives

$$\langle 0 \| \psi(\Delta t) \rangle = \frac{1}{(i\hbar)^2} \frac{e^2}{c^2} \int_0^{\Delta t} dt' \int_0^{t'} dt'' \langle 0 |\Big[ \int d^3\mathbf{r}' \,\hat{\mathbf{j}}(\mathbf{r}',t') \cdot \hat{\mathbf{A}}(\mathbf{r}',t') \Big]$$
$$\times \Big[ \int d^3\mathbf{r}'' \,\hat{\mathbf{j}}(\mathbf{r}'',t'') \cdot \hat{\mathbf{A}}(\mathbf{r}'',t'') \Big] \| E_1 \rangle |G_2\rangle |0\rangle. \tag{A1}$$

From Eq. (7) in the text, the matrix element $M$ of the field operators is given by

$$M \equiv \left\langle 0 \left| \hat{A}_x(\mathbf{r}',t') \hat{A}_x(\mathbf{r}'',t'') \right| 0 \right\rangle = -ic\hbar D_F(\mathbf{r}'-\mathbf{r}'',t'-t'')$$

$$= \frac{c\hbar}{4\pi^2} \frac{1}{|\mathbf{r}'-\mathbf{r}''|^2 - c^2(t'-t'')^2}. \tag{A2}$$

We can expand this in a Taylor series about the point $p$ given by $\mathbf{r}' = \mathbf{x}_2$, $\mathbf{r}'' = \mathbf{x}_1$, $t' = 0$, and $t'' = 0$:

$$M = \frac{c\hbar}{4\pi^2} \left\{ \frac{1}{r^2} + \left[ \frac{\partial f}{\partial x'} \right]_p (x'-x_2) + \left[ \frac{\partial f}{\partial y'} \right]_p (y'-y_2) + ... \right.$$

$$\left. + \left[ \frac{\partial f}{\partial t'} \right]_p t' + \left[ \frac{\partial f}{\partial t''} \right]_p t'' \right\} \tag{A3}$$

Here we have defined $r = |\mathbf{x}_2 - \mathbf{x}_1|$ as before and the function $f$ is defined as

$$f = \frac{1}{|\mathbf{r}'-\mathbf{r}''|^2 - c^2(t'-t'')^2}. \tag{A4}$$

The partial derivatives are given by

$$\left[ \frac{\partial f}{\partial x'} \right]_p = -\left[ \frac{2(x'-x'')}{\left( |\mathbf{r}'-\mathbf{r}''|^2 - c^2(t'-t'')^2 \right)^2} \right]_p = -\frac{2}{r^4}(x_2 - x_1)$$

$$\left[ \frac{\partial f}{\partial t'} \right]_p = \left[ \frac{2c^2(t'-t'')}{\left( |\mathbf{r}'-\mathbf{r}''|^2 - c^2(t'-t'')^2 \right)^2} \right]_p = 0 \tag{A5}$$

for example, with similar expressions for the other terms. For simplicity, we can take $\mathbf{x}_2 - \mathbf{x}_1$ along the z axis, so that only the partial derivatives with respect to $z'$ and $z''$ are nonzero.

Inserting this into Eq. (A1) and taking the projection onto $\left| G_1 \right\rangle \left| E_2 \right\rangle$ gives the probability amplitude $b$:

$$b = \frac{1}{(i\hbar)^2} \frac{e^2}{c^2} \frac{c\hbar}{4\pi} \frac{1}{r^2} \int_0^{\Delta t} dt' \int_0^{t'} dt''$$

$$\times \left\langle E_2 \left| \left[ \int d^3\mathbf{r}' \, \hat{j}_x(\mathbf{r}',t') \left( 1 - \frac{2}{r}(z'-z_2) + ... \right) \right] \right| G_2 \right\rangle$$

$$\times \left\langle G_1 \left| \left[ \int d^3\mathbf{r}'' \, \hat{j}_x(\mathbf{r}'',t'') \left( 1 + \frac{2}{r}(z''-z_1) + ... \right) \right] \right| E_1 \right\rangle. \tag{A6}$$

It can be seen that the second term in the expansion is smaller than the first by a factor of $d_A / r$, where $d_A$ is the characteristic dimensions of the atom. Thus in the limit of $d_A / r \ll 1$ we get

$$b = \frac{1}{(i\hbar)^2} \frac{e^2}{c^2} \frac{c\hbar}{4\pi} \frac{1}{r^2} \int_0^{\Delta t} dt' \int_0^{t'} dt'' \left\langle E_2 \left| \left[ \int d^3\mathbf{r}' \, \hat{j}_x(\mathbf{r}',t') \right] \right| G_2 \right\rangle$$
$$\times \left\langle G_1 \left| \left[ \int d^3\mathbf{r}'' \, \hat{j}_x(\mathbf{r}'',t'') \right] \right| E_1 \right\rangle.$$

(A7)

The identity $\left\langle j_x \right\rangle = -iE_A d_x / \hbar$ does not depend on the dipole approximation and can be derived from the fact that $dx/dt = \hat{p}_x / m = \left[ x, \hat{H}_0 \right] / i\hbar$ [36]. Eq. (A7) is thus equivalent to Eqs. (9) and (10) in the text.

These results show that the contributions from the higher-order multipoles are negligible when the separation between the atoms is much larger than their dimensions. This situation is very different from the usual case in which electromagnetic energy is emitted by atom 1 and then travels at the speed of light to atom 2, where it can be absorbed; there the ratio of the dipole to quadrupole contributions is independent of the distance, for example. In that case the largest contribution is on the light cone where $D_F$ diverges and the Taylor's series expansion is not valid. Here we are considering the opposite limit, where the multipole contributions can be completely neglected.

## APPENDIX B: DRESSED ATOMIC STATES AND VIRTUAL PHOTONS

The analysis in the text assumed an initial state in which atom 1 was in its bare excited state $\left| E_1 \right\rangle$ and atom 2 was in its bare ground state $\left| G_2 \right\rangle$ with no photons present initially. It is valid to consider the time evolution of an initial state of this kind, at least as a Gedanken experiment, and most of the previous discussion has been based on this example. It is shown here that similar results are obtained in the more realistic case in which the initial state corresponds to dressed atomic states that have some probability amplitude to include one or more virtual photons. The entanglement generated during the time interval $\Delta t$ will be shown to be independent of any virtual photons that existed before that time to lowest order in perturbation theory.

The presence of divergent diagrams, such as the self-corrections to the mass and charge of the electron, require that the theory be renormalized. The calculations can then be performed using the physical mass and charge, as is implicitly done in most quantum optics calculations. It will be assumed that the theory has been renormalized in this way, so that the remaining interaction is small ($\alpha \ll 1$) and perturbation theory can be used.

After the renormalization, the remaining interaction of the atoms with the field will produce energy eigenstates with some probability amplitude for the presence of a virtual photon, which we will refer to as dressed states. The virtual photon "cloud" associated with a dressed atomic state is illustrated in a very schematic way by the black arrows in Fig. 7a. We will refer to an eigenstate of the Hamiltonian that does not include the interaction with the field (but with the theory renormalized to use the physical mass and charge of the electron) as a bare atomic state.

Including dressed atomic states in the analysis raises a number of questions: Does the presence of the virtual photons in the initial dressed states affect the matrix elements and the transition amplitudes? Is the entanglement between dressed or bare states, and how would that affect the results of any measurements? And would the virtual photons in the final dressed states have any effect on the outcome of the detectors used in the post-selection process?

In order to investigate these issues, we will make two basic assumptions: (i) The interaction between the field and the atoms is weak ( $\alpha \ll 1$ ) after renormalization, so that we can consider only the lowest-order effects. (ii) The interaction Hamiltonian that couples the atomic states of interest ( $\left| E \right\rangle$ and $\left| G \right\rangle$ ) can be effectively turned off before time $t = 0$ and after time $t = \Delta t$ , or at least reduced to a negligible value outside the time interval $\Delta t$ .

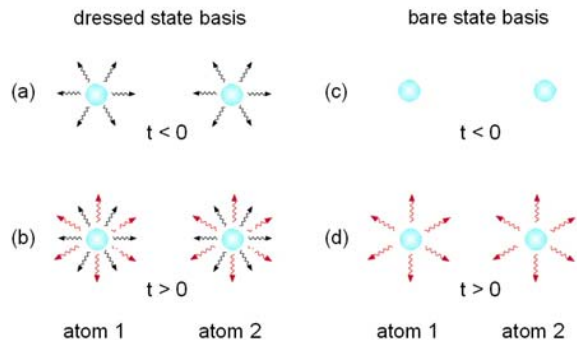

FIG.7. Comparison of the dressed state basis with the bare state basis. (a) After renormalization, the dressed atomic states will include virtual photons (represented by black arrows) produced by atomic transitions other than the one of interest for times $t < 0$ . (b) During the time interval $\Delta t$ , the interaction between the ground state and first excited state is turned on by an external field, which produces additional virtual photons represented by red arrows. (c) The effects of the additional photons are independent of the virtual photons that existed before $t = 0$ to lowest order, so that the situation is equivalent to using the bare initial state with no virtual photons. (d) During the time interval $\Delta t$ , the analysis in the bare state basis includes only the virtual photons generated by the atomic transition of interest. To lowest order, this gives the same results as the dressed state analysis of (a) and (b).

As mentioned in the text, an example of a situation in which the interaction Hamiltonian can be modulated is the case in which the excited dressed states $\left| E_1 ' \right\rangle$ and $\left| E_2 ' \right\rangle$ correspond to metastable states with zero dipole moment. Atom 1 can then be assumed to have been in dressed state $\left| E_1 ' \right\rangle$ for times $t < 0$ and similarly for atom 2 in dressed state $\left| G_2 ' \right\rangle$ . The application of an external electric field $\mathbf{E}_{ext}$ can be used to produce a dipole moment for both atoms over the time interval $\Delta t$ , as in the Lamb shift experiments. We will work in the interaction picture where the unperturbed Hamiltonian $\hat{H}_0(t)$ is chosen to include the effects of $\mathbf{E}_{ext}$ and the coupling to the field that exists for $t < 0$ , while the interaction Hamiltonian $\hat{H}'(t)$ will include only the coupling term that is turned on during time $\Delta t$ .

Modulating the interaction in this way provides a clear definition of the initial conditions that apply to the problem. Moreover, the modulation could be applied using electromagnetic pulses traveling in free space, so that there is no change in the boundary conditions associated with the modulation. This avoids the ambiguities that have been associated with some of the earlier discussions. It should also be noted that the virtual photons that exist prior to $t = 0$ all correspond to atomic transitions other than the $|G\rangle \leftrightarrow |E\rangle$ transition of interest, which has zero matrix elements for $t < 0$ .

We begin with a generalization of Eqs. (13-15) in the text, which showed that $|\psi\rangle \neq |\Psi_1\rangle|\Psi_2\rangle$ in the bare state basis. Here the dressed atomic states may contain correlations between the two atoms even before $t < 0$ , and the bare atomic states used in the main text need to be replaced with more general eigenstates, such as

$$|E_1\rangle|G_2\rangle|0\rangle \rightarrow |E_1',G_2',0'\rangle. \tag{B1}$$

Here $|E_1',G_2',0'\rangle$ corresponds to the perturbed eigenstate produced by the interaction with the field in $\hat{H}_0$ , which will no longer factor into two single-atom states as before. The $0'$ in Eq. (B1) reflects the fact that the field will no longer be in the bare vacuum state and it may contain virtual photons. Similar notation will be used for the other dressed states of interest.

The dressed atomic states that exist outside of the time interval $\Delta t$ are eigenstates of $\hat{H}_0$ . To lowest order in perturbation theory, they can be written in the form

$$\begin{aligned}
|E_1',G_2',0'\rangle &= \beta|E_1\rangle|G_2\rangle|0\rangle + \varepsilon\sum_{\mathbf{k}i}c(\mathbf{k},i)|F_{1i}\rangle|G_2\rangle|\mathbf{k}\rangle \\
&+ \varepsilon\sum_{\mathbf{k}i}d(\mathbf{k},i)|E_1\rangle|F_{2i}\rangle|\mathbf{k}\rangle + \varepsilon^2\sum_{\mathbf{k}\mathbf{p}ij}e(\mathbf{k},\mathbf{p},i,j)|F_{1i}\rangle|F_{2j}\rangle|\mathbf{k},\mathbf{p}\rangle \\
&+ \varepsilon^2\sum_{ij}f(i,j)|F_{1i}\rangle|F_{2j}\rangle|0\rangle + \varepsilon^2 g|G_1\rangle|G_2\rangle|0\rangle \\
&+ \varepsilon^2 h|E_1\rangle|E_2\rangle|0\rangle
\end{aligned} \tag{B2}$$

for example, with similar expressions for the other dressed states. Here $\varepsilon$ is a constant of the same order as $\hat{H}'$ , $\beta$ is a normalization constant on the order of unity, $c$ through $h$ represent arbitrary complex coefficients, and $|F_{1i}\rangle$ and $|F_{2i}\rangle$ represent atomic states other than the ground state or first excited state. The form of this equation follows from the fact that $\hat{H}_0$ does not directly couple the ground and first excited states.

The second and third terms in Eq. (B2) correspond to the emission of a single virtual photon and an atomic transition to $|F_{1i}\rangle$ or $|F_{2i}\rangle$ . The fourth term corresponds to the emission of two virtual photons and two atomic transitions, while the fifth term corresponds to the emission of a virtual photon by one atom followed by its absorption by the other atom. The final two terms correspond to the emission and absorption of a virtual photon by the same atom, which can produce a transition from $|E_1\rangle$ to $|G_1\rangle$ , for example, via a virtual state involving $|F_{1i}\rangle$ . (Similar transitions to $|F_{1i}\rangle|G_2\rangle|0\rangle$ and $|E_1\rangle|F_{2i}\rangle|0\rangle$ are of no interest and have not been included.) It should be noted that the term $|G_1\rangle|E_2\rangle|0\rangle$ can only be produced via two virtual transitions of that kind and is therefore fourth order and negligible.

We will consider the subspace S of Hilbert space corresponding to the projection $\hat{P}_{EG}$ onto the ground and first excited atomic states:

$$\hat{P}_{EG} \equiv |G_1\rangle|G_2\rangle\langle G_2|\langle G_1| + |G_1\rangle|E_2\rangle\langle E_2|\langle G_1| +$$
$$|E_1\rangle|G_2\rangle\langle G_2|\langle E_1| + |E_1\rangle|E_2\rangle\langle E_2|\langle E_1|. \tag{B3}$$

The projection of the initial dressed state onto S gives

$$\hat{P}_{EG}|E_1',G_2',0'\rangle = \beta|E_1\rangle|G_2\rangle|0\rangle + \varepsilon^2 g|G_1\rangle|G_2\rangle|0\rangle$$
$$+\varepsilon^2 h|E_1\rangle|E_2\rangle|0\rangle. \tag{B4}$$

This state can be prepared by local operations on the two atoms and it can be written as the product of two single-atom states to order $\varepsilon^2$. This shows that the initial state is not entangled in the subspace S before the interaction at time $t = 0$ to lowest order.

We will now show that the system is entangled in subspace S after the interaction over time interval $\Delta t$. The Hamiltonian $\hat{H}_0$ cannot produce any change $|\Delta\psi\rangle$ in the state of the system and $|\Delta\psi\rangle$ must involve at least one factor of $\hat{H}'$. To second order in $\varepsilon$, we can therefore drop the $\varepsilon^2$ terms in Eq. (B2) and only the first three terms in the initial state can contribute to $|\Delta\psi\rangle$. Since $\hat{H}'$ does not produce any transitions involving the states $|F_{1i}\rangle$, the projection onto the subspace S also has no contribution from the second and third terms of Eq. (B2), so that

$$\hat{P}_{EG}|\Delta\psi\rangle = \frac{1}{(i\hbar)^2}\int_0^{\Delta t} dt'\int_0^{t'} dt''\hat{H}'(t')\hat{H}'(t'')|E_1\rangle|G_2\rangle|0\rangle. \tag{B5}$$

where we have taken $\beta \to 1$. This is identical to Eq. (2) in the main text, and the same techniques used there can be used to show that

$$\hat{P}_{EG}|\psi\rangle \neq |\Psi_1\rangle|\Psi_2\rangle \tag{B6}$$

for $t > \Delta t$.

These results show that the system was not entangled in subspace S before the interaction (to second order) while it is entangled in subspace S after the interaction. It is also apparent that the entanglement that is generated during time interval $\Delta t$ is independent of any entanglement or virtual photons that may have existed prior to the interaction to lowest order.

The situation here is analogous to entanglement in quantum optics, where a photon can be independently entangled in polarization or in energy-time variables, for example. The entanglement in polarization is routinely measured experimentally while simply ignoring any energy-time entanglement. Here the entanglement generated during the interaction over time interval $\Delta t$ is orthogonal to any entanglement that previously existed, and the latter can be ignored in the same way.

We will now consider the effects of using dressed atomic states for post-selection and entanglement concentration, as in Fig. 3. The analysis can be repeated using the same perturbation theory techniques that were used in the main text or by using Eq. (17), where the matrix elements will now be taken in the dressed-state basis. A transition from the dressed state $|E_1',G_2',0'\rangle$ to $|G_1',E_2',0'\rangle$ will require two factors of $\hat{H}'$ as described above, so that a typical term $T$ in the integrand is

$$T = \left\langle G_1 ' E_2 ' 0 \right| \hat{H}\,' \left| G_1 ' G_2 '\mathbf{k} \right\rangle \left\langle G_1 ' G_2 '\mathbf{k} \right| \hat{H}\,' \left| E_1 ' G_2 ' 0 \right\rangle \tag{B7}$$

for example. Since the factors of $\hat{H}\,'$ already introduce a factor of $\varepsilon^2$, to second order in perturbation theory this reduces to

$$\begin{aligned} T &= \Big[ \big( \langle 0 | \langle E_2 | \langle G_1 | \big) \hat{H}\,' \big( |G_1\rangle |G_2\rangle |\mathbf{k}\rangle \big) \Big] \\ &\times \Big[ \big( \langle \mathbf{k} | \langle G_2 | \langle G_1 | \big) \hat{H}\,' \big( |E_1\rangle |G_2\rangle |0\rangle \big) \Big]. \end{aligned} \tag{B8}$$

These matrix elements are the same as in the bare-state basis, and the perturbation calculation will therefore give the same results as were obtained in the main text to lowest order. Measuring the joint state of the two systems is a nonlocal operation [54, 55] that will require a longer time interval $\Delta T$ as before, but that does not alter the basic conclusion that entanglement can be generated outside of the forward light cone.

We can model the detectors as additional two-level atoms initially prepared in their dressed ground states, where a detection event will correspond to finding the atom in its excited dressed state. We do not require a fast response from the detectors, so that they can be turned on slowly after time $\Delta t$ and the system allowed to evolve over a relatively long time period during the post-selection process (but still shorter than $r/c$). As a result, energy conservation will apply. For times $t < 0$, the system is in an eigenstate of the Hamiltonian where no transitions can occur and the detectors could not have registered a count due to any virtual photons in the initial dressed atomic states. The same situation will hold after time $\Delta t$, where atoms 1 and 2 are either in their ground state or a metastable state and cannot supply any energy to excite the detector atoms. As a result, the detector atoms cannot respond to any virtual photons associated with the dressed atomic eigenstates and they can only register photons that were emitted by atom 1 or atom 2 during the time interval $\Delta t$. Thus the post-selection process can be performed as described in the main text.

Since we are working in the dressed state basis, Eq. (B8) combined with the post-selection and entanglement concentration described in the main text will produce a final state of the form

$$\begin{aligned} |\psi''\rangle &= \Big[ \left| E_1 ', G_2 ', 0 \right\rangle + \left| G_1 ', E_2 ', 0 \right\rangle \Big] / \sqrt{2} \\ &= \Big[ |E_1\rangle |G_2\rangle |0\rangle + |G_1\rangle |E_2\rangle |0\rangle \Big] / \sqrt{2}. \end{aligned} \tag{B9}$$

The second line of the equation holds to second order in the interaction from Eq. (B2), since the dominant terms are already second order. Thus the dressed-state analysis gives the same results as the bare-state analysis in the text to lowest order.

A comparison of the dressed-state and bare-state analyses is summarized in Fig. 7. In the dressed-state basis, there will be virtual photons associated with other atomic transitions for $t < 0$, as illustrated by the black arrows in Fig. 7a. Turning on the coupling between the ground and first excited states over time interval $\Delta t$ will produce additional virtual photons, as illustrated by the red arrows in Fig. 7b. The effects of these additional virtual photons are independent of the virtual photons that were present in the initial state to lowest order, so that the net result is the same as in the bare-state basis illustrated in Figs. 7c and 7d. This justifies the use of the bare-state basis in the main text and in the discussion of Appendix C.

## APPENDIX C:  COVARIANT CALCULATION

It seems apparent that these results are due to the nature of the Feynman propagator for the photons and that the atoms can just as well be treated nonrelativistically.  It has been suggested, however, that the calculations should be performed in a covariant way nevertheless to ensure that the results are not an artifact of the nonrelativistic treatment of the atoms.  It will be shown in this appendix that the same results are obtained using the second-quantized Dirac theory for the electrons, aside from a small relativistic correction to the atomic matrix elements.  In the Lorentz gauge, the interaction Hamiltonian can be written in the form

$$\hat{H}' = -\int d^3\mathbf{r}\, \hat{j}_\mu \hat{A}^\mu \tag{C1}$$

Here  $\hat{j}_\mu$  is the current 4-vector in the second-quantized Dirac theory with components

$$\hat{\rho}(\mathbf{r}) = q\hat{\psi}^\dagger(\mathbf{r})\hat{\psi}(\mathbf{r})$$
$$\hat{\mathbf{j}}(\mathbf{r}) = qc\hat{\psi}^\dagger(\mathbf{r})\boldsymbol{\alpha}\hat{\psi}(\mathbf{r}) \tag{C2}$$

where $\hat{\psi}(\mathbf{r})$ is the field operator for the electron-positron field and $\boldsymbol{\alpha}$  corresponds to the Dirac matrices.  The use of perturbation theory will give rise to integrals  $I$  of the form

$$I = \int_{-\infty}^{t} \hat{H}'(t')dt' = -\int_{-\infty}^{t} dt' \int d^3\mathbf{r}\, \hat{j}_\mu \hat{A}^\mu \tag{C3}$$

for example.  If the interaction goes to zero after the time of interest, which it does in S-matrix theory as well as in the metastable state example discussed above, then the integral over time can be extended to infinity to give

$$I = -\int_{-\infty}^{\infty} d^4x\, \hat{j}_\mu \hat{A}^\mu \tag{C4}$$

Eq. (C4) is an invariant under Lorentz transformations, and it can be shown that the results of perturbation theory are the same in any reference frame and that perturbation theory is equivalent to the use of Wick's theorem in scattering (S-matrix) calculations [35].

In the Lorentz gauge, the interaction Hamiltonian includes both the vector and scalar potential operators:

$$\hat{H}' = \int d^3\mathbf{r}(-\hat{\mathbf{j}}\cdot\hat{\mathbf{A}} + \hat{\rho}\hat{\Phi}) \tag{C5}$$

Here $\hat{\Phi}$  is the scalar potential operator and  $\hat{\rho}$  is the charge density operator.  The contribution from the scalar potential term can be evaluated using the commutator techniques of Appendix A. That contribution vanishes in the limit of  $r \gg c\Delta t$ , since the only nonzero matrix elements of  $\hat{\rho}$ depend on the partial derivatives of Eq. (A5).  Thus the only significant contribution is from the vector potential term that was used in Eq. (3) of the main text.

Although the nonrelativistic Schrodinger equation was tacitly assumed in the main text, the actual form of the current operator was never used. Instead, the dipole approximation was derived from the identity

$$dx/dt = \hat{p}_x/m = \left[ x, \hat{H}_0 \right]/i\hbar \qquad (C6)$$

This relationship must hold at least approximately for the Dirac theory, since the Schrodinger equation corresponds to its nonrelativistic limit. As a result, the matrix elements of $\hat{j}_i$ in the Dirac theory are related at least approximately to the dipole moment by $-iE_A d_i/\hbar$, just as in the nonrelativistic Schrodinger equation. The dipole moment is to be evaluated between the relativistic eigenstates of the atom [56], which will give a small correction to their value.

Since the matrix elements of the interaction Hamiltonian are the same in the Dirac theory as in the nonrelativistic Schrodinger equation, aside from a small relativistic correction to the dipole moments, it follows from perturbation theory or Eq. (17) that the Dirac theory will give all the same results that were previously described in the main text. This discussion also shows that the same results would be observed in any reference frame, as can be explicitly demonstrated.

Hegerfeldt [52] showed that, in relativistic quantum field theory, a free particle that is completely localized inside a finite volume at time $t = 0$ will subsequently have some probability to be found arbitrarily far away after a short time interval. It has been argued [13, 54, 55] that this difficulty in localizing free particles invalidates the usual assumption that the initial state is localized. But Hegerfeldt's theorem applies to free particles, whereas the electrons in the atomic states of interest here are bound. In addition, they are not strictly localized in the initial state, since the relativistic atomic state corresponds to exponentially decaying probability amplitudes [56], and Hegerfeldt's theorem does not apply. To within an exponentially small error, the atomic eigenstates can be modulated by an external electric field of finite extent and the effects of that modulation outside of the forward light cone can be determined as discussed in the text.